\documentclass[aps,twocolumn,superscriptaddress]{revtex4}
\usepackage{amsmath}
\usepackage{amsfonts}
\usepackage{mathrsfs}
\usepackage{graphicx}
\usepackage{times}
\usepackage{color}
\usepackage[normalem]{ulem}

\newcommand\redout{\bgroup\markoverwith{\textcolor{red}{\rule[.5ex]{2pt}{1pt}}}\ULon}

\def\be{\begin{equation}}
\def\ee{\end{equation}}
\def\bea{\begin{eqnarray}}
\def\eea{\end{eqnarray}}

\begin{document}

\title{Manipulating Goldstone modes via the superradiant light in a bosonic lattice gas inside a cavity}

\author{Huan Wang}
\affiliation{MOE Key Laboratory for Nonequilibrium Synthesis and Modulation of Condensed Matter,Shaanxi Province Key Laboratory of Quantum Information and Quantum Optoelectronic Devices, School of Physics, Xi'an Jiaotong University, Xi'an 710049, China}

\author{Shuai Li}
\affiliation{MOE Key Laboratory for Nonequilibrium Synthesis and Modulation of Condensed Matter,Shaanxi Province Key Laboratory of Quantum Information and Quantum Optoelectronic Devices, School of Physics, Xi'an Jiaotong University, Xi'an 710049, China}

\author{Maksims Arzamasovs}
\affiliation{MOE Key Laboratory for Nonequilibrium Synthesis and Modulation of Condensed Matter,Shaanxi Province Key Laboratory of Quantum Information and Quantum Optoelectronic Devices, School of Physics, Xi'an Jiaotong University, Xi'an 710049, China}

\author{W. Vincent Liu}
\thanks{On academic leave from Department of Physics and Astronomy, University of Pittsburgh, Pittsburgh PA 15260, USA}
\affiliation{Department of Physics and Shenzhen Institute for Quantum Science and Engineering, Southern University of Science and Technology, Shenzhen, 518055, China}

\author{Bo Liu}
\email{liubophy@gmail.com}
\affiliation{MOE Key Laboratory for Nonequilibrium Synthesis and Modulation of Condensed Matter,Shaanxi Province Key Laboratory of Quantum Information and Quantum Optoelectronic Devices, School of Physics, Xi'an Jiaotong University, Xi'an 710049, China}

\begin{abstract}
We study the low-energy excitations of a bosonic lattice gas with cavity-mediated interactions. By performing two successive Hubbard-Stratonovich transformations, we derive an effective field theory to study the strongly-coupling regime. Taking into account the quantum fluctuation, we report the unusual effect of the superradiant cavity light induced density imbalance, which has been shown to have a negligible effect on the single particle excitation in the previous studies. Instead, we show that such negligible fluctuation of density imbalance dramatically changes the behavior of the low-energy excitation and results in a free switching between two types of Goldstone modes in its single particle excitation, i.e., type I and type II with odd and even power energy-momentum dispersion, respectively. Our proposal would open a new horizon for manipulating Goldstone modes from bridging the cavity light and strongly interacting quantum matters.
\end{abstract}

\maketitle

%\section{Introduction}
The mechanism of spontaneous symmetry breaking is crucial for understanding phase transitions and is also widely used to study the associated emergence of new particles and excitations~\cite{1987_Huang}. It is well known that when a continuous symmetry is spontaneously broken in nonrelativistic theories, there appear Nambu-Goldstone (NG) modes~\cite{2_1960_Nambu,3_1961_Goldstone}. The dispersion relations of which are either linear (type I) or quadratic (type II), where the numbers of distinct types of NG modes satisfy the Nielsen-Chadha inequality~\cite{4_1976_Nielsen}. Thanks to recent experimental developments, the Goldstone modes have been studied in various condensed matter~\cite{5_1930_Bloch,6_1994_Chubukov,7_2012_Podolsky,8_2007_Ye} and ultra-cold atomic systems ~\cite{9_2010_Ernst,10_2008_Papp,11_2002_Steinhauer,12_1999_Stenger,13_1999_Stamper,14_1999_Kozuma,2013_Yi,PhysRevLett_vincent}.

A gas of bosonic atoms in an optical lattice has been reversibly tuned between superfluids (SF) and insulating ground states by varying the strength of the periodic potential~\cite{15_2002_Greiner,16_2004_Stoferle}. It provides an ideal platform to study the spontaneous symmetry breaking induced elementary excitations. Not only has the gapless Goldstone mode been found to exhaust all of the spectral weight in the weakly-interacting limit, {but also the Higgs amplitude mode has been detected in the strongly-interacting regime
near the transition between SF and Mott state at commensurate fillings~\cite{17_2011_Sachdev,18_2012_Endres,19_2015_Pekker}}.

Recently, a complimentary approach using ultracold atoms inside a cavity unveils the effect of cavity-mediated interactions, which results in the observation of a {rich phase diagram} with Mott insulator, superfluid, supersolid (SS) and charge-density-wave (CDW) phases~\cite{20_2016_Landig,21_2015_Klinder}. The low-energy excitations associated with above various symmetry-breaking phases have been investigated within the framework of mean-field approach, where new features, such as the softening of particle- and hole-like excitations, have been explored~\cite{22_2016_Dogra}. The quantum fluctuation has not been taken into account within such mean-field calculations. However, it has been shown to play an important role in the strongly-interacting regime, leading to intriguing physical phenomena, such as the appearance of Higgs mode in the excitation spectrum of the Bose-Hubbard model~\cite{23_1994_Freericks,24_1996_Freericks}.

In this work, a systematic strong-coupling expansion method has been developed by performing two successive Hubbard-Stratonovich transformations~\cite{25_2005_Sengupta} for the system of ultracold bosons in an optical lattice strongly coupled to a single mode of a high-finesse optical cavity~\cite{20_2016_Landig,21_2015_Klinder}.  The effect of quantum fluctuations can thus been systematically explored in the strongly-interacting regime, which has not been studied previously~\cite{22_2016_Dogra,26_2016_Chen}. {Interestingly, we find that the quantum fluctuations at the SS-to-CDW transition cause an unexpected effect of the modulation of density imbalance, neglected in the previous studies ~\cite{22_2016_Dogra}. While being small, this modulation results in dramatic changes of the behavior of low-energy excitations. A free switching between the odd (type I) and even (type II) power energy-momentum dispersion NG modes can be achieved along the phase boundary. This prediction would pave a new way to manipulate NG modes in a cold atom based system.}

%\section{Effective model}
\textit{Effective model $\raisebox{0.01mm}{---}$} {Same as the ETH experiment~\cite{20_2016_Landig},
here we consider a Bose-Einstein condensate (BEC), such as $^{87}$Rb atoms, subjected to an ultrahigh-finesse optical cavity, where three mutually orthogonal standing waves
form a highly anisotropic $3$D optical lattice $V_{L}(\mathbf{r})=V_{y}\cos^{2}(q_{0}y)+V_{2D}[\cos^{2}(q_{0}x)+\cos^{2}(q_{0}z)])$
with the lattice depth $V_{y}\gg V_{2D}$ and the wave vectors of
laser fields denoted by $q_0$}. The BEC is thus split into a stack of $2$D layers,
where the bosonic atoms are exposed to an additional potential $V_{c}(\mathbf{r})=\hbar\bar{\eta} (\hat{a}+\hat{a}^{\dag}) \cos(q_{0}x)\cos(q_{0}z)-\hbar(\Delta_{c}-U_{0}\cos^{2}(q_{0}z))\hat{a}\hat{a}^{\dag} $
in the $xz$-plane formed by the coherent scattering between one free space lattice and one intracavity optical standing wave with the same wave-vector $q_0$~\cite{20_2016_Landig,21_2015_Klinder}.
$\bar{\eta}$ is the two-photon Rabi frequency determining the scattering rate.
$\hat{a}$ and $\hat{a}^{\dag }$ are the annihilation and creation operators for the cavity photon, respectively. $\Delta_{c}=\omega_{p}-\omega_{c}$ describes the discrepancy between pumping light (lattice standing-wave) frequency $\omega_{p}$ and the cavity resonance frequency $\omega_{c}$. $U_{0}$ captures the maximum light shift per atom resulting from the effect of the dispersive shift of the cavity resonance frequency~\cite{20_2016_Landig}.

Since the coherent scattering of light between the
lattice and the cavity mode creates a dynamical checkerboard superlattice for the
atoms, the effective Hamiltonian describing the atomic dynamics dressed by the cavity field can be expressed as
\begin{eqnarray}
H&=&\sum_{\mathbf{r}, \mathbf{r}^{\prime}}-\mathrm{T}_{\sigma \sigma^{\prime}}\left(\mathbf{r}-\mathbf{r}^{\prime}\right) \psi_{\sigma}^{\dagger}(\mathbf{r}) \psi_{\sigma^{\prime}}\left(\mathbf{r}^{\prime}\right)-\sum_{\mathbf{r},\sigma} \mu_{\sigma} n_{\sigma}(\mathbf{r}) \nonumber \\
&+&\frac{U}{2} \sum_{\mathbf{r},\sigma} n_{\sigma}(\mathbf{r})
(n_{\sigma}(\mathbf{r})-1)-\delta_{c}|\alpha|^{2},
\label{2}
\end{eqnarray}
where $\psi_{\sigma}(\mathbf{r})$ is the bosonic atom field.
$\sigma=e, o$ stands for even and odd sites, respectively, describing distinct
sublattices of the checkerboard lattice. Here we choose the nearest
neighbor two sites as one unit cell and $\mathbf{r}$ ($\mathbf{r}^{\prime}$) is
the lattice index capturing the location of the unit cell. The expression of the hopping matrix $\mathrm{T}_{\sigma \sigma^{\prime}}$ is given in the supplementary material (SM). $U$ captures the strength of the repulsive interaction between bosonic atoms determined by the effective $s$-wave scattering length, which can be tuned by means of the Feshbach resonance and lattice depth. $n_{\sigma}(\mathbf{r})=\psi_{\sigma}^{\dagger}(\mathbf{r}) \psi_{\sigma}(\mathbf{r})$ is the density operator. The onsite energies $\mu_{e}=\mu-\eta\left(\alpha^{*}+\alpha\right)$ and $\mu_{o}=\mu+\eta\left(\alpha^{*}+\alpha\right)$ are introduced for even and odd sites, respectively, where $\mu$ is the
chemical potential and $\eta$ is the energy shift due to the two-photon Rabi process. $\delta_c=\hbar (\Delta_c-\delta)$ is the energy detuning between the cavity and the pumping
lights, where $\delta$ is the dispersive shift of the cavity due to the BEC~\cite{20_2016_Landig}. {$\alpha$ is the mean value of the cavity field, which can be determined by the steady equation
$i\hbar\partial_{t}\alpha=<[\hat{\alpha},{\tilde{\mathbf H}}]>-i\kappa\alpha=0$
with ${\tilde{\mathbf H}}=-\sum_{\mathbf{r}, \mathbf{r}^{\prime}}\mathrm{T}_{\sigma \sigma^{\prime}}\left(\mathbf{r}-\mathbf{r}^{\prime}\right) \psi_{\sigma}^{\dagger}(\mathbf{r}) \psi_{\sigma^{\prime}}\left(\mathbf{r}^{\prime}\right)-\mu \sum_{\mathbf{r},\sigma} n_{\sigma}(\mathbf{r}) +\frac{U}{2} \sum_{\mathbf{r},\sigma} n_{\sigma}(\mathbf{r})
(n_{\sigma}(\mathbf{r})-1)+\eta (\hat{a}+\hat{a}^{\dag})\sum_{\mathbf{r}}(n_{e}(\mathbf{r})-n_{o}(\mathbf{r}))-\hbar (\Delta_c-\delta)\hat{a}^{\dag}\hat{a}$
in the regime of large decay rate $\kappa$~\cite{Baumann2010} }and leads to the following relation
\begin{eqnarray}
\alpha &&=\frac{\eta \sum_{\mathbf{r}}\left[<n_{e} (\mathbf{r})>-
<n_{o}(\mathbf{r})> \right]}{\delta_{c}+i\kappa} \nonumber \\
&&\overset{\left\vert \Delta _{c}\right\vert \gg \kappa ,\left\vert
\delta\right\vert }{\approx }\frac{N\eta \left( \left\langle
n_{e}\right\rangle -\left\langle n_{o}\right\rangle \right) }{\hbar \Delta _{c}}
\label{add1}
\end{eqnarray}
where $<n_{e} (\mathbf{r})>$ and $<n_{o}(\mathbf{r})>$ are the average density of bosonic atoms at even and odd sites, respectively. $2N$ is the total lattice site.

\begin{figure*}
\begin{center}
\includegraphics[width=0.9\textwidth]{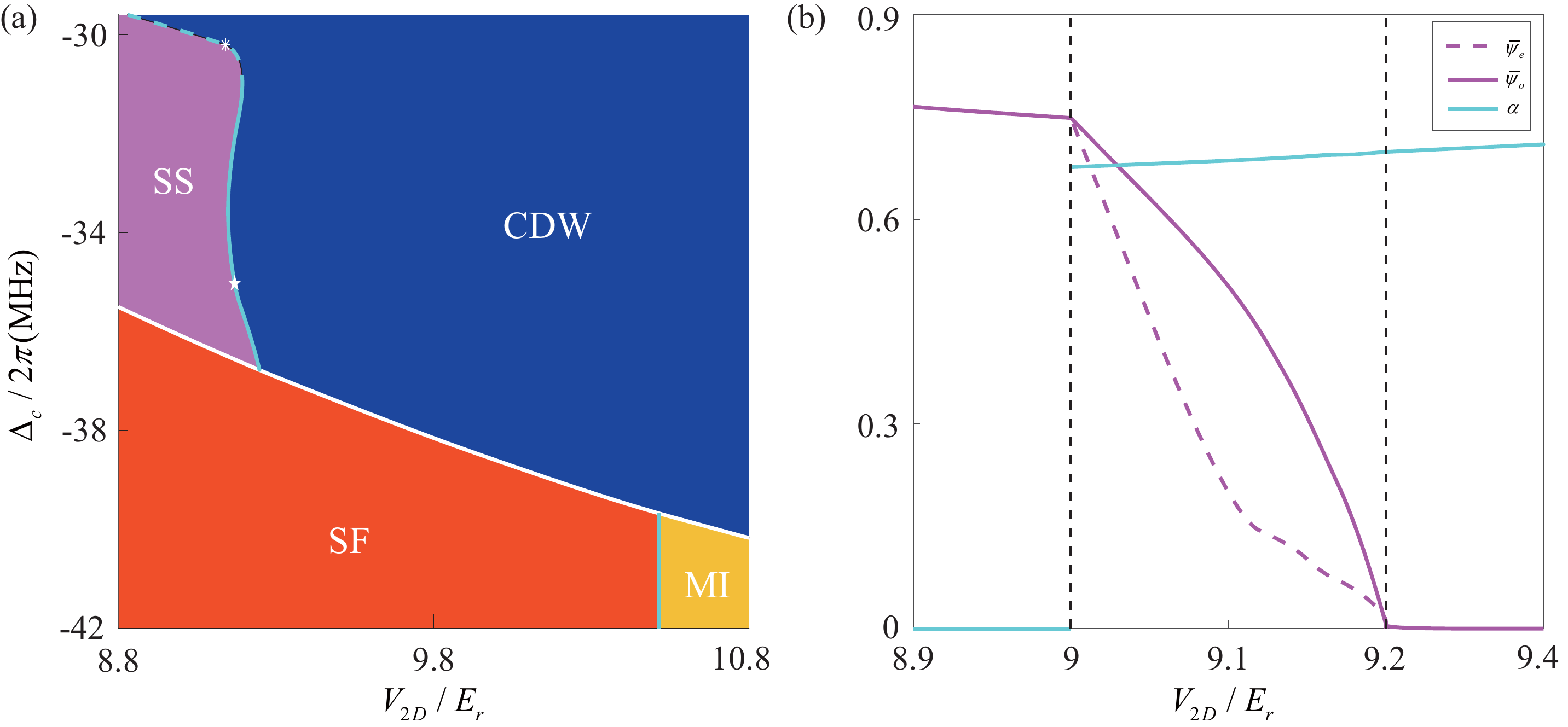}
\end{center}
\caption{(a) Zero temperature phase diagram as a function
of the lattice depth $V_{2D}$ and $\Delta_{c}$. There are four different phases
as shown in the phase diagram, which consists of a superfluid (SF) phase, a Mott insulator
(MI) state, a charge-density-wave (CDW)state and a supersolid (SS) phase.
The dashed and solid line separate the two distinct region, where two different types of Goldstone modes (type-I and type-II) were found in its single particle excitation, respectively. (b) Evolution of the different order parameters as a function of the lattice depth $V_{2D}$ when $\Delta_{c}/2\pi=-36$MHz. $E_r$ is the recoil energy. Other parameters are chosen as  $\mu=0.42U$, $\kappa=2\pi \times 1.25$MHz, $N=1 \times 10^4$.
\label{fig:wide}}
\end{figure*}

%\section{Path integral approach}
\textit{Path integral approach $\raisebox{0.01mm}{---}$} {In the following, we develop a strong-coupling expansion of the model Hamiltonian in Eq.~\eqref{2},
which extends the treatment in Bose-Hubbard model~\cite{25_2005_Sengupta}, where both superfluid and Mott phases can be captured in the same time by this method.} Through introducing the complex field $\psi_{\sigma}$, the thermodynamic properties of the system can be obtained from the partition function $Z$ as a functional integral
with the action $S\left[\psi_{\sigma}^{*}, \psi_{\sigma}\right]=\int_{0}^{\beta} d \tau\left\{\sum_{\sigma, \mathbf{r}} \psi_{\sigma}^{*}(\mathbf{r}) \partial_{\tau} \psi_{\sigma}(\mathbf{r})+H\left[\psi^{*}, \psi\right]\right\}$.
Here, $\tau$ is a imaginary time and $\beta = 1/k_BT$ is inverse temperature.
We then perform a Hubbard-Stratonovich transformation through introducing
the auxiliary field $\phi_{\sigma}$ to decouple the intersite hopping
in the action and obtain
\begin{eqnarray}
&&Z =\int D\left[\psi_{\sigma}^{*}, \psi_{\sigma}, \phi_{\sigma}^{*}, \phi_{\sigma}\right] \exp\{-\int_{0}^{\beta} d \tau \sum_{\mathbf{r}, \mathbf{r}^{\prime}}\phi_{\sigma}^{*}(\mathbf{r}) T_{\sigma \sigma^{\prime}}^{-1} \phi_{\sigma^{\prime}}(\mathbf{r'}) \nonumber \\
&&+[\sum_{\sigma, \mathrm{r}} \int_{0}^{\beta} d \tau \phi_{\sigma}^{*}(\mathbf{r}) \psi_{\sigma}(\mathbf{r})+c . c]-S_{0}\left[\psi_{\sigma}^{*}, \psi_{\sigma}\right]\}\nonumber \\
&&= Z_{0} \int D\left[\phi_{\sigma}^{*}, \phi_{\sigma}\right] e^{-\int_{0}^{\beta} d \tau \sum_{\mathbf{r}, \mathbf{r}^{\prime}}\phi_{\sigma}^{*}(\mathbf{r})  T_{\sigma \sigma^{\prime}}^{-1} \phi_{\sigma^{\prime}}(\mathbf{r'})}\nonumber \\
&&\times\langle\exp [\sum_{\sigma, \mathrm{r}} \int_{0}^{\beta} d \tau \phi_{\sigma}^{*}(\mathbf{r}) \psi_{\sigma}(\mathbf{r})+c . c]\rangle_{0} \nonumber \\
&&= Z_{0} \int D\left[\phi_{\sigma}^{*}, \phi_{\sigma}\right] \exp \{-\int_{0}^{\beta} d \tau \sum_{\mathbf{r}, \mathbf{r}^{\prime}}\phi_{\sigma}^{*} T_{\sigma \sigma^{\prime}}^{-1} \phi_{\sigma^{\prime}}+W\left[\phi_{\sigma}^{*}, \phi_{\sigma}\right]\},\nonumber \\
\label{5}
\end{eqnarray}
where $T_{\sigma \sigma^{\prime}}^{-1}$ represents the inverse hopping matrix. $S_{0}$ and $Z_{0}$ are the action and partition function in the limit of $t=0$. $\langle\cdots\rangle_{0}$ stands for averaging with $S_{0}$. $W\left[\phi_{\sigma}^{*}, \phi_{\sigma}\right]=\ln \langle\exp [\sum_{\sigma, \mathrm{r}} \int_{0}^{\beta} d \tau \phi_{\sigma}^{*}(\mathbf{r}) \psi_{\sigma}(\mathbf{r})+\mathrm{c.c.}]\rangle_{0}$ is the generation function
linking to connected local Green's functions $G^{R c}$ through the relation [25] $W\left[\phi^{*}, \phi\right]=\sum_{R=1}^{\infty} \frac{(-1)^{R}}{(R !)^{2}} \sum'_{\sigma_{1} \cdots \sigma_{R}^{\prime}} G_{\left\{\sigma_{i}, \sigma_{i}^{\prime}\right\}}^{R c} \phi_{\sigma_{1}}^{*} \cdots \phi_{\sigma_{R}}^{*} \phi_{\sigma_{R}^{\prime}} \cdots \phi_{\sigma_{1}^{\prime}}$,
where $\sum'$ means that all the fields share the same value of the
site index and ${\left\{\sigma_{i}, \sigma_{i}^{\prime}\right\}}\equiv {\sigma_{1}\cdots\sigma_{R},\sigma_{1}^{\prime}\cdots\sigma^{\prime}_{R}}$. After doing a power expansion of $W\left[\phi^{*} , \phi\right]$ to quartic order, we obtain that
\begin{eqnarray}
S[\phi_{\sigma}^{*},\phi_{\sigma}]&=&\int_{0}^{\beta} d \tau \sum_{\mathbf{r}, \mathbf{r}^{\prime}} \phi_{\sigma}^{*}(\mathbf{r}) T_{\sigma \sigma^{\prime}}^{-1}\left(\mathbf{r}-\mathbf{r}^{\prime}\right) \phi_{\sigma^{\prime}}\left(\mathbf{r}^{\prime}\right)-W\left[\phi_{\sigma}^{*}, \phi_{\sigma}\right] \nonumber \\
&=&\int_{0}^{\beta} d \tau \sum_{\mathbf{r}, \mathbf{r}^{\prime}} \phi_{\sigma}^{*}(\mathbf{r}) T_{\sigma \sigma^{\prime}}^{-1}\left(\mathbf{r}-\mathbf{r}^{\prime}\right) \phi_{\sigma^{\prime}}\left(\mathbf{r}^{\prime}\right)\nonumber \\
&+&\int d\tau _{1}d\tau _{2}\underset{\mathbf{r}}{\sum }G_{\sigma \sigma
}(\mathbf{r,}\tau _{1}-\tau _{2}\mathbf{)}\phi _{\sigma }^{\ast }(\mathbf{r}%
,\tau _{1})\phi _{\sigma }(\mathbf{r},\tau _{2}) \nonumber \\
&-&\frac{1}{2!}\int \underset{\alpha =1}{\overset{4}{\Pi }}d\tau _{\alpha }%
\underset{\mathbf{r}}{\sum }\chi _{\sigma \sigma ^{\prime }}(\mathbf{r,}\tau
_{1},\tau _{2},\tau _{3},\tau _{4})\nonumber \\
&\times& \phi _{\sigma }^{\ast }(\tau _{1})\phi _{\sigma }(\tau _{2})\phi
_{\sigma ^{\prime }}^{\ast }(\tau _{3})\phi _{\sigma ^{\prime }}(\tau
_{4})+O(\phi ^{6}), \label{7}
\end{eqnarray}
where $G$ is the local Green's function and $\chi$ is the two-particle vertex in
the local limit, i.e., $t=0$ (see details in SM). However, starting from the above action, it is inconvenient to calculate physical quantities, such as
the excitation spectrum or the momentum distribution. These difficulties can be circumvented through performing another Hubbard-Stratonovich transition to decouple
the hopping term~\cite{25_2005_Sengupta}. It is shown that the auxiliary field of this transformation
has the same correlation functions as the original boson field (see details in SM). Therefore, we can use the same notation for both fields. The partition function can thus
be expressed as
\begin{eqnarray}
&&Z=Z_{0} \int D\left[\psi_{\sigma}^{*}, \psi_{\sigma}, \phi_{\sigma}^{*}, \phi_{\sigma}\right] \exp \{\int_{0}^{\beta} d \tau \sum_{\mathbf{r}, \mathbf{r}^{\prime}}\psi_{\sigma}^{*} T_{\sigma \sigma^{\prime}} \psi_{\sigma^{\prime}}\nonumber \\
&&-[\sum_{\sigma, \mathrm{r}} \int_{0}^{\beta} d \tau \psi_{\sigma}^{*}(\mathbf{r}) \phi_{\sigma}(\mathbf{r})
+ \mathrm{c.c.}]+W\left[\phi_{\sigma}^{*}, \phi_{\sigma}\right]\} ,
\label{8}
\end{eqnarray}
Integrating out the $\phi_{\sigma}$ field in Eq.~\eqref{8}, the effective action can be obtained
\begin{eqnarray}
&&S[\psi_{\sigma}^{*},\psi_{\sigma}]= \int_{0}^{\beta} d \tau \sum_{\mathbf{r}, r^{\prime}} \psi_{\sigma}^{*}(\mathbf{r}, \tau)\left[-T_{\sigma \sigma^{\prime}}\left(\mathbf{r}-\mathbf{r}^{\prime}\right)\right] \psi_{\sigma^{\prime}}\left(\mathbf{r}^{\prime}, \tau\right) \nonumber \\
&&+\int_{0}^{\beta} d \tau_{1} d \tau_{2} \sum_{\mathbf{r}} \psi_{\sigma}^{*}\left(\mathbf{r}, \tau_{1}\right)\left[-G_{\sigma \sigma}^{-1}\left(\tau_{1}-\tau_{2}\right)\right] \psi_{\sigma}\left(\mathbf{r}, \tau_{2}\right)\nonumber \\
&&+\frac{1}{2} g_{\sigma \sigma^{\prime}} \int_{0}^{\beta} d \tau \sum_{\mathbf{r}}\left|\psi_{\sigma}(\mathbf{r}, \tau)\right|^{2}\left|\psi_{\sigma^{\prime}}(\mathbf{r}, \tau)\right|^{2},
\label{9}
\end{eqnarray}
where $g_{\sigma \sigma^{\prime}}$ captures the amplitude of the boson-boson
interaction determined by the local two-particle vertex in its static limit (see details in SM).

\begin{figure}[t]
\includegraphics[width=8cm]{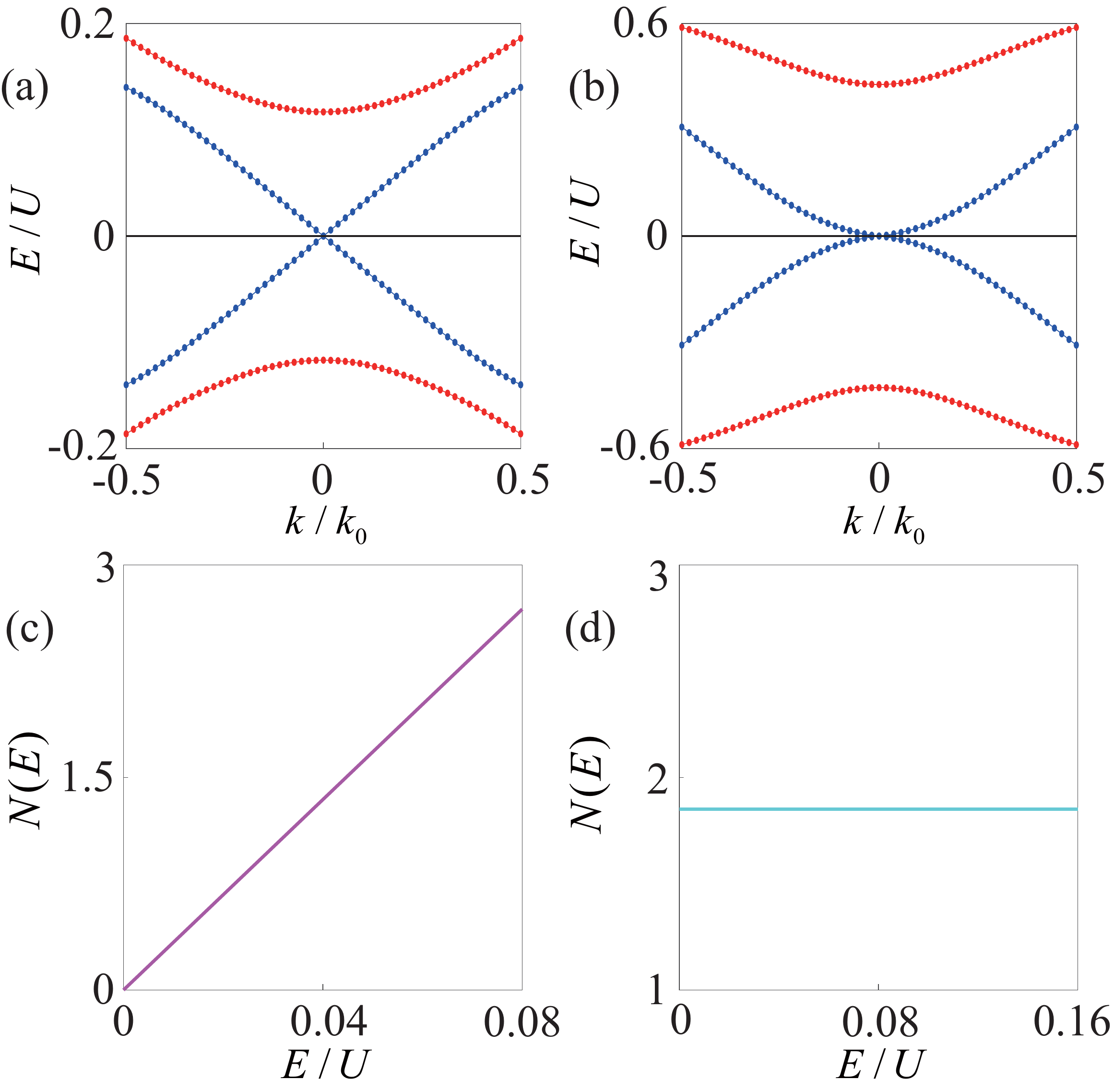}
\caption{ Excitation spectra for the two lowest-energy excitations (a) when
$V_{2D}/E_{r}=9.1$, $\Delta_{c}/2\pi=-30.2$MHz as indicated by '$\ast$' in Fig. 1(a), (b) when $V_{2D}/E_{r}=9.14$, $\Delta_{c}/2\pi=-35$MHz as labelled by '$\star$' in Fig. 1(a). Other parameters are the same as Fig. 1. (c) (d) Density of states (DOS) defined in the main text for type I and type II Goldstone modes, respectively.
\label{fig:GM}}
\end{figure}

%\section{Phase diagram under the saddle-point approximation}
\textit{Phase diagram under the saddle-point approximation $\raisebox{0.01mm}{---}$}
Starting from the above effective action, we first perform a saddle-point approximation
to determine the ground state of our proposed system. The saddle-point action derived from Eq.~\eqref{9} can be expressed as
\begin{eqnarray}
S_{saddle} &=& - \bar G_{ee}^{ - 1}{\left|  {{{\bar\psi}_{e}}} \right| ^{2}} - \bar
G_{oo}^{ - 1}{\left|  {{{\bar\psi}_{o}}} \right| ^{2}} - 4t{{\bar\psi}_{o}%
}^{*}{{\bar\psi}_{e}} - 4t{{\bar\psi}_{e}}^{*}{{\bar\psi}_{o}}\nonumber \\
&-& \frac{1}{4}\frac{{{{\bar\chi}_{ee}}}}{{\bar G_{ee}^{4}}}{\left|  {{{\bar\psi}_{e}}}
\right| ^{4}}
- \frac{1}{4}\frac{{{{\bar\chi}_{oo}}}}{{\bar G_{oo}^{4}}%
}{\left|  {{{\bar\psi}_{o}}} \right| ^{4}},
\label{91}
\end{eqnarray}
where $\bar{G}_{\sigma\sigma}\equiv G_{\sigma\sigma}(i\omega=0)$ and $\bar{\chi}_{\sigma\sigma}\equiv \chi_{\sigma\sigma}(i\omega=0)$ with $G_{\sigma\sigma}(i\omega)$ and $\chi_{\sigma\sigma}(i\omega)$ being
Fourier transforms of the single-particle Green's function
and the two-particle vertex, respectively. The saddle-point value $\bar{\psi}_{\sigma}$ can be obtained from
minimizing $S_{saddle}$ and we obtain
\begin{eqnarray}
n_{s,e}&=&\frac{\bar{G}_{ee}^{-1}+4tc^{-1}}{g_{ee}} \text{ \ if }\bar{G}_{ee}^{-1}+4tc^{-1}>0 \nonumber \\
n_{s,e}&=&0\text{ \ \ \ \ \ \ \ \ \ \ \ \ \ \ \ \ \ \ \ \ \ \ \ \ \ \ \ \ \ otherwisde,} \\
n_{s,o}&=&\frac{\bar{G}_{oo}^{-1}+4tc}{g_{oo}}  \text{ \ \ \ \ \ \ if }\bar{G}_{oo}^{-1}+4tc>0 \nonumber \\
n_{s,o}&=&0\text{ \ \ \ \ \ \ \ \ \ \ \ \ \ \ \ \ \ \ \ \ \ \ \ \ \ \ \ \ \ \ otherwise}, \label{14}
\end{eqnarray}
where $n_{s,o(e)}=|\bar{\psi}_{\sigma}|^2$ is the superfluid density of the pseudospin ${\sigma}$ component and $c=\bar{\psi}_{e}/\bar{\psi}_{o}$. By utilizing an iterative algorithm to solve a complete set of self-consistent equations (2), (8) and (9), superfluid order parameters $\bar{\psi}_{\sigma}$ for even and odd lattice sites and the mean value of cavity field $\alpha$ can be determined. As defined in Eq. (2), $\alpha$ characterizes the average imbalance between even and odd lattice sites, which can be used as the order parameter describing the CDW order.

We display the phase diagram as a function of the lattice depth $V_{2D}$ and detuning $\Delta_{c}$ in Fig.~\ref{fig:wide}(a), to compare with the realistic experiments, such as the ETH experimental setup~\cite{20_2016_Landig}. There are four different phases as shown in the phase
diagram, which consists of a superfluid (SF) phase, a Mott
insulator (MI) state, a charge-density-wave (CDW) state and a supersolid (SS) phase.
The difference among these four phases can be captured by order parameters defined
above, for example, as shown in  Fig.~\ref{fig:wide}(b). A SF phase is characterized by
finite and equal superfluid order parameters and vanishing even-odd imbalance indicated by $\alpha=0$. It is quite different in the SS phase, where a finite even-odd imbalance and nonzero superfluid order parameters are observed. In both MI and CDW states,   superfluid order parameters vanish. However, the presence of a finite even-odd imbalance
$\alpha \neq 0$ distinguishes {between} MI and CDW states. The obtained phase diagram is consistent with other mean-field calculations~\cite{22_2016_Dogra,26_2016_Chen}. However, the path integral approach constructed here can provide a systematic
way beyond the mean-field approach for investigating the effect of high-order fluctuations. The unexpected results
driven by the fluctuations will be unveiled below.

\begin{figure}[t]
\includegraphics[width=8cm]{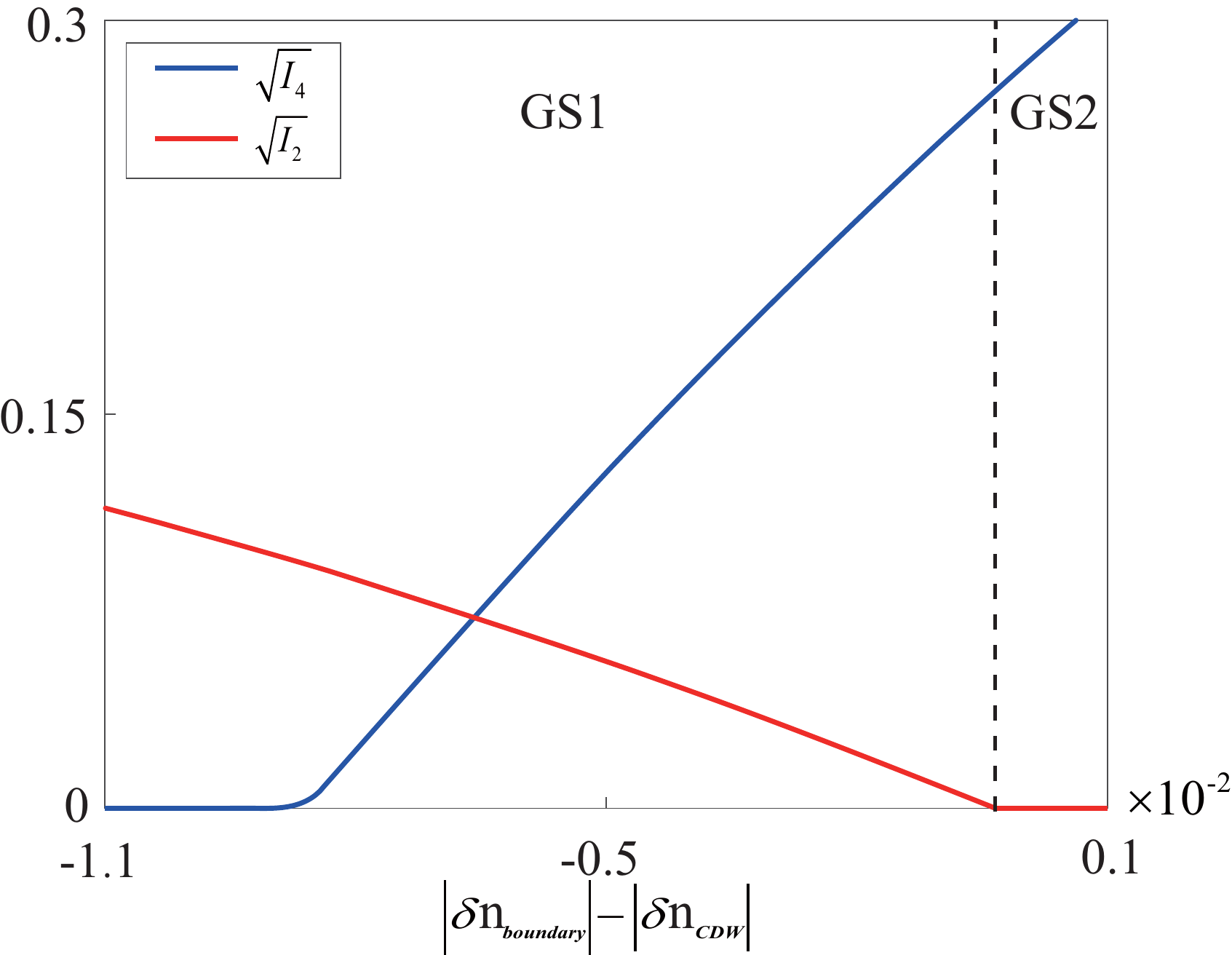}
\caption{{The coefficients $I_2$ and $I_4$ in $E^{+}_{Goldstone}$ (Eq.~\ref{GNA}) as a function of density imbalance modulation when varying the system  along the phase boundary between SS and CDW as shown in Fig. 1(a).} The dashed line separates the regions with
type I and type II Goldstone modes, respectively. $\delta n_{boundary}$ and $\delta n_{CDW}$ describe the atom density imbalance between even and odd sites, when considering the system along the phase boundary and in CDW, respectively. Here $\delta n_{CDW}$ is fixed at $\delta n_{CDW}=2$. Other parameters are chosen as the same in Fig. 1.\label{fig:co}}
\end{figure}

%\section{Manipulating Goldstone modes at the SS-CDW transition}
\textit{Manipulating Goldstone modes at the SS-CDW transition $\raisebox{0.01mm}{---}$}
Since typically fluctuations will play a dominate role in determining physical  properties at the critical region, to explore the unexpected results arising from
the fluctuations, we construct the critical theory describing the phase transition region as shown in Fig.~\ref{fig:wide}(a). By performing both a spatial and a temporal gradient expansion of the action in Eq. (6), as well as the cumulant expansion in powers of
$\psi_{\sigma}$~\cite{25_2005_Sengupta,PhysRevB_Fisher,sym10040080}, the effective action can be rewritten as
\begin{widetext}
\begin{eqnarray}
\begin{aligned}
S =&\int\nolimits_{0}^{\beta }d\tau \int d\mathbf{r\{}\left[ \psi
_{e}^{\ast }(\mathbf{r},\tau )r_{e}\psi _{e}(\mathbf{r},\tau )+\psi
_{o}^{\ast }(\mathbf{r},\tau )r_{e}\psi _{o}(\mathbf{r},\tau )-\psi
_{o}^{\ast }(\mathbf{r},\tau )4t\psi _{e}(\mathbf{r},\tau )-\psi _{e}^{\ast
}(\mathbf{r},\tau )4t\psi _{o}(\mathbf{r},\tau )\right]   \label{15} \\
&+\left[ K_{e1}\psi _{e}^{\ast }(\mathbf{r},\tau )\partial _{\tau }\psi
_{e}(\mathbf{r},\tau )+K_{o1}\psi _{o}^{\ast }(\mathbf{r},\tau )\partial
_{\tau }\psi _{o}(\mathbf{r},\tau )\right] +\left[ K_{e2}\left\vert \partial
_{\tau }\psi _{e}(\mathbf{r},\tau )\right\vert ^{2}+K_{o2}\left\vert
\partial _{\tau }\psi _{o}(\mathbf{r},\tau )\right\vert ^{2}\right]  \\
&+\left[ K_{3}\nabla \psi _{e}^{\ast }(\mathbf{r},\tau )\nabla \psi _{o}(%
\mathbf{r},\tau )+K_{3}\nabla \psi _{o}^{\ast }(\mathbf{r},\tau )\nabla \psi
_{e}(\mathbf{r},\tau )\right] +\frac{g_{ee}}{2}\left\vert \psi _{e}(\mathbf{r}%
,\tau )\right\vert ^{4}+\frac{g_{oo}}{2}\left\vert \psi _{o}(\mathbf{r},\tau
)\right\vert ^{4} \\
&+\frac{g_{eo}}{2}\left\vert \psi _{e}(\mathbf{r},\tau )\right\vert
^{2}\left\vert \psi _{o}(\mathbf{r},\tau )\right\vert ^{2}+\frac{g_{oe}}{2}%
\left\vert \psi _{o}(\mathbf{r},\tau )\right\vert ^{2}\left\vert \psi _{e}(%
\mathbf{r},\tau )\right\vert ^{2}\}, \label{17}
\end{aligned}
\end{eqnarray}
\end{widetext}
where $r_{e}=-{G}_{ee}^{-1}(i\omega=0)$, $r_{o}=-{G}_{oo}^{-1}(i\omega=0)$,
$K_{e1}=\frac{\partial G_{ee}^{-1}\left( i\omega\right) }{\partial
(i\omega)}|_{i\omega=0}$, $K_{o1}=\frac{\partial G_{oo}^{-1}\left(
i\omega\right) }{\partial (i\omega )}|_{i\omega=0}$.
$K_{e(o)2}$ and $K_{3}$ are the coefficients of the second-order temporal and spatial derivatives, respectively, which can be expressed in terms of the Green's function (see
SM for details).

Next, we introduce $\eta_{\sigma}(\mathbf{r},\tau )=\psi_{\sigma}(\mathbf{r},\tau )-\bar{\psi} _{\sigma}$ to describe the fluctuations. After expanding the action in Eq. (10) to the quadric order of fluctuation fields $\eta_{\sigma}$, we get
\begin{eqnarray}
S=\frac{1}{2}\underset{\mathbf{k},\omega}{\sum }\tilde{\eta}^{\dag}({\mathbf{k},i\omega })M({\mathbf{k},i\omega})\tilde{\eta}({\mathbf{k},i\omega}),
\label{19}
\end{eqnarray}
where $\tilde{\eta}^{\dag}=[\eta _{e}^{\ast }(\mathbf{k},i\omega ), \eta _{e}(\mathbf{-k},-i\omega ), \eta_{o}^{\ast}(\mathbf{k},i\omega ),$ $\eta _{o}(\mathbf{-k},-i\omega )]$ and the matrix elements of $M$ can be expressed as

\begin{eqnarray}
\begin{split}
&M_{11} = r_{e}-i\omega K_{e1}+K_{e2}\omega ^{2}+2g_{ee}\bar{\psi} _{e}^{2}+\frac{%
g_{eo}}{2}\bar{\psi} _{o}^{2}+\frac{g_{oe}}{2}\bar{\psi} _{o}^{2}, \nonumber \\
&M_{22} =r_{e}+i\omega K_{e1}+K_{e2}\omega ^{2}+2g_{ee}\bar{\psi} _{e}^{2}+\frac{%
g_{eo}}{2}\bar{\psi} _{o}^{2}+\frac{g_{oe}}{2}\bar{\psi} _{o}^{2}, \nonumber \\
&M_{33} =r_{o}-i\omega K_{o1}+K_{o2}\omega ^{2}+2g_{oo}\bar{\psi} _{o}^{2}+\frac{%
g_{eo}}{2}\bar{\psi} _{e}^{2}+\frac{g_{oe}}{2}\bar{\psi} _{e}^{2}, \nonumber \\
&M_{44} =r_{o}+i\omega K_{o1}+K_{o2}\omega ^{2}+2g_{oo}\bar{\psi} _{o}^{2}+\frac{%
g_{eo}}{2}\bar{\psi} _{e}^{2}+\frac{g_{oe}}{2}\bar{\psi} _{e}^{2}, \nonumber \\
&M_{12} =M_{21}=g_{ee}\bar{\psi} _{e}^{2},\quad M_{34}=M_{43}=g_{oo}\bar{\psi} _{o}^{2}, \nonumber \\
&M_{13} =M_{24}=M_{31}=M_{42} \nonumber \\
&=-4t+K_{3}k^{2}+\frac{g_{eo}}{2}\bar{\psi}
_{e}\bar{\psi} _{o} +\frac{g_{oe}}{2}\bar{\psi} _{o}\bar{\psi} _{e}, \nonumber \\
&M_{14} =M_{23}=M_{32}=M_{41}=\frac{g_{eo}}{2}\bar{\psi} _{e}\bar{\psi} _{o}+\frac{%
g_{oe}}{2}\bar{\psi} _{o}\bar{\psi} _{e}. \label{20}
\end{split}
\end{eqnarray}
Therefore, the single particle excitation spectrum at the critical region as shown in Fig. 1(a), can be obtained by solving
the equation $det[M({\mathbf{k},i\omega})]=0$.

Here we focus on the transition from SS to CDW. In SS phase, there are two superfluid order parameters $\bar{\psi}_{\sigma}\equiv \sqrt{n_{s,\sigma}}e^{i\theta_{\sigma}}$ for even and odd lattice sites, where the amplitude and phase of the order parameter emerge as two independent degree of freedoms, instead of being conjugate to each other. The phase fluctuation leads to the Goldstone mode, while the amplitude fluctuation leads to the Higgs mode. When further considering the coupling between even and odd lattice sites (hopping term between them), the two Higgs modes arising from the amplitude fluctuation will couple to each other and split into two new Higgs modes (two higher modes obtained from $det[M({\mathbf{k},i\omega})]=0$). While the two Goldstone modes resulting from the phase fluctuation will also couple to each other and split into one gapless Goldstone mode associated with the overall phase $\theta_{o}+\theta_{e}$ plus a gapped mode linking to the relative phase $\theta_{o}-\theta_{e}$.

At the transition from SS to CDW, we find the new feature of the lowest Goldstone mode excitation. Previous studies show that the slight modulation of  density imbalance induced by the superradiant cavity light has a negligible effect on the single particle excitation spectrum~\cite{22_2016_Dogra}. Here we firstly show that at the transition from SS to CDW, even such negligible modulation of density imbalance will result in a dramatic effect on the excitations. Distinct from the mean-field approach~\cite{22_2016_Dogra,26_2016_Chen}, such as employing
the Gutzwiller ansatz, the critical theory constructed in Eq.~\eqref{17} can unveil the
dominate role of fluctuations. It is found that tuning the system along the phase boundary between SS and CDW as shown in Fig~\ref{fig:wide}(a) can freely switch between
two types of Goldstone modes in its single particle excitation, i.e., type I and type II with odd and even power energy-momentum dispersion (as shown in Fig.~\ref{fig:GM}), respectively.

To further understand the underlying physics, we analytically solve the equation $det[M({\mathbf{k},i\omega})]=0$ and find that in the long wave limit the lowest single particle excitation (positive branch) can be approximately expressed as (See details in SM)
\begin{eqnarray}
E^{+}_{Goldstone}\simeq\sqrt{I_{2}}k+\sqrt{I_{4}}k^{2}.
\label{GNA}
\end{eqnarray}
Therefore, the switching from type I to type II Goldstone modes can be achieved by changing the coefficients $I_{2}$ and $I_{4}$. Around the transition from SS to CDW, it is shown that the quantum fluctuations make even the slight modulation of density imbalance play a dominant role in determining $I_{2}$ and $I_{4}$. As shown in Fig.~\ref{fig:co},
the slight modulation of density imbalance leads to two distinct regions: (I) $I_{2}\neq 0$, (II)  $I_{2} = 0$ and $I_{4}\neq 0$, which corresponds type I and type II Goldstone modes, respectively. To distinguish these two different types of Goldstone modes, we calculate the density of states (DOS) for the lowest single particle excitation
as $N(E)=\frac{1}{N}\sum_{\mathbf k}\delta(E-E^{+}_{Goldstone})$, which is directly related to the Bragg spectroscopy signal~\cite{2009PhysRevA}. As shown in Fig.~\ref{fig:GM}, with linear dispersion (type I), we find $N(E) \propto E$ when $E\rightarrow 0$. It is distinct from the quadratic dispersion (type II), where N(E) is a constant when $E\rightarrow 0$. The experimental advances in the momentum-resolved Bragg spectroscopy in optical lattices~\cite{Ernst2010} make the detection of this signal accessible.

%\section{Conclusion}
\textit{Conclusion $\raisebox{0.01mm}{---}$}
By performing two successive Hubbard-Stratonovich transformations, we have
developed an effective field theory to study a bosonic lattice gas inside
a cavity in the strongly interacting regime. Through taking into account the quantum
fluctuations, we firstly find that the slight modulation of density imbalance, neglected in the previous studies, leads to dramatic changes of the behavior of low-energy excitations. The switching between type I and type II Goldstone modes can be driven along SS-CDW transition. Our findings would bridge the cavity light and strongly interacting quantum matters and may open up a new direction in this field.

\textit{Acknowledgement $\raisebox{0.01mm}{---}$}
This work is supported by NSFC (Grant No.
12074305, 11774282, 11950410491), the National Key Research
and Development Program of China (2018YFA0307600), Cyrus Tang Foundation Young Scholar
Program and the Fundamental Research Funds for the Central Universities(H.W., S.L., M.A and B.L.) and by the AFOSR Grant No. FA9550-16-1-0006, the MURI-ARO Grant No. W911NF17-1-0323 through UC Santa Barbara, and the Shanghai Municipal Science and Technology Major Project through the Shanghai Research Center for Quantum Sciences (Grant No. 2019SHZDZX01) (W.V.L.). We also thank the HPC platform of Xi'An Jiaotong University, where our numerical calculations was performed.

\bibliographystyle{apsrev}
\bibliography{GM}

\onecolumngrid

%%%%%%%%%%%%%%%%%%%%%%%%%%%%%%%%%%%%%%
%%   Supplementary Information
%%%%%%%%%%%%%%%%%%%%%%%%%%%%%%%%%%%%%%
%\appendix
\renewcommand{\thesection}{S-\arabic{section}}
\setcounter{section}{0}  %  this will re-count section from 1
\renewcommand{\theequation}{S\arabic{equation}}
\setcounter{equation}{0}  %  this will re-count eq from 1
\renewcommand{\thefigure}{S\arabic{figure}}
\setcounter{figure}{0}  %  this will re-count eq from 1

\indent

\begin{center}\large
\textbf{Supplementary Material:\\ Manipulating Goldstone modes via the superradiant light in a bosonic lattice gas inside a cavity}
\end{center}

\section{Hopping term}
\begin{figure}[tbh]
\includegraphics[width=6cm]{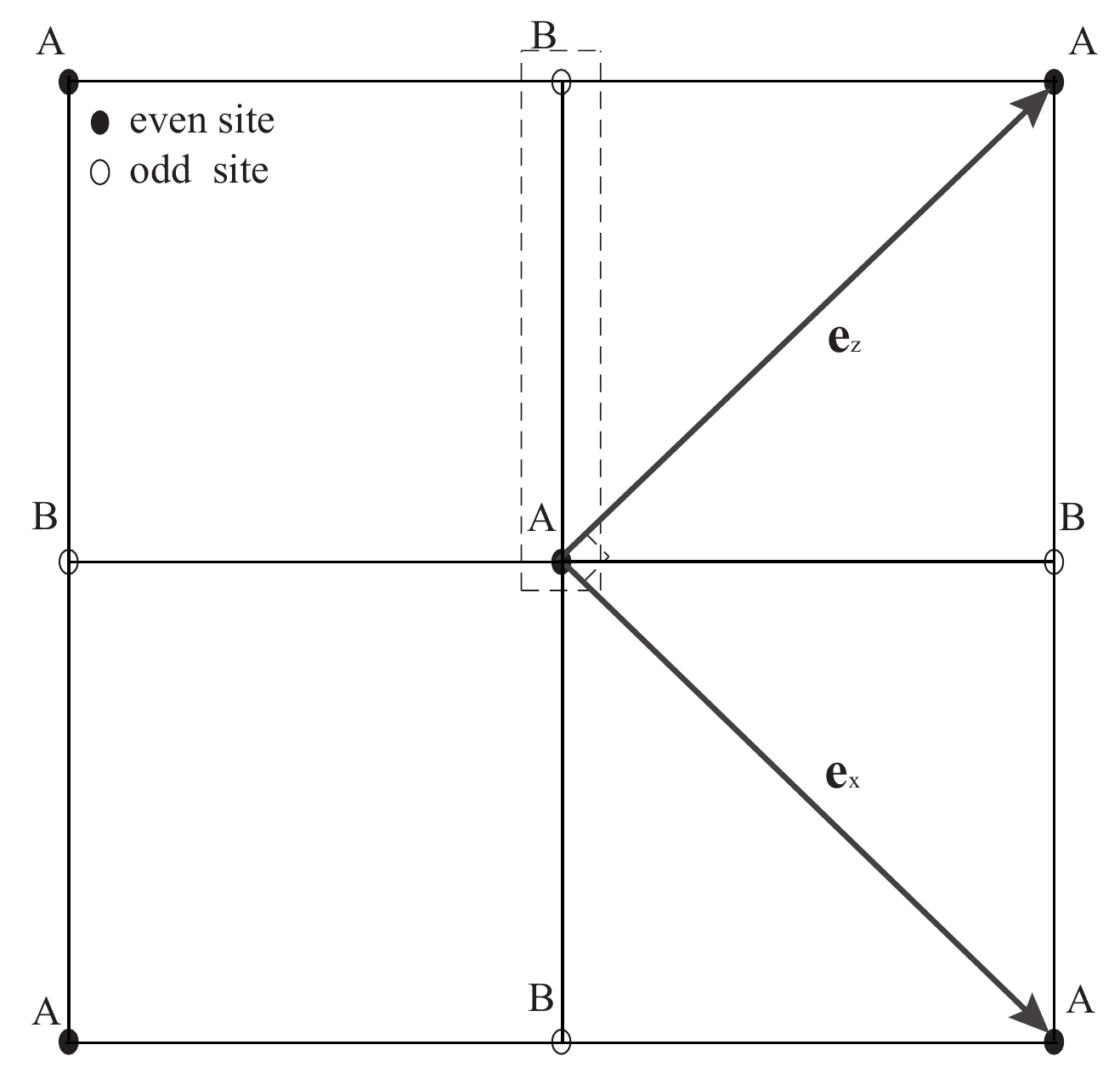}
\caption{Schematic picture of a two dimensional checkerboard lattice. Here A and B
stand for two different sites in one unit cell(even and odd sites), ${\mathbf{e_x}}$ and ${\mathbf{e_z}}$ are the primitive unit vectors.}
\label{fig:lattice}
\end{figure}

The hopping term in Eq. (1) can be written as
\begin{eqnarray}
H_{hop}&=&\underset{\mathbf{r},\mathbf{r}^{\prime }}{\sum }-T_{\sigma \sigma
^{\prime }}(\mathbf{r}-\mathbf{r}^{\prime })\psi _{\sigma }^{\dag }(\mathbf{r%
})\psi _{\sigma ^{\prime }}(\mathbf{r'})\notag \\
&=&\underset{\mathbf{r}}{\sum }\psi _{\sigma }^{\dag }(\mathbf{r}%
)T_{0}\psi _{\sigma }(\mathbf{r})+\psi _{\sigma }^{\dag }(\mathbf{r})T_{1x}\psi _{\sigma }(\mathbf{r+e}%
_{x})+\psi _{\sigma }^{\dag }(\mathbf{r})T_{1x}^{\prime }\psi _{\sigma }(%
\mathbf{r-e}_{x})  \notag \\
&+&\psi _{\sigma }^{\dag }(\mathbf{r})T_{1z}\psi _{\sigma }(\mathbf{r+e}%
_{z})+\psi _{\sigma }^{\dag }(\mathbf{r})T_{1z}^{\prime }\psi _{\sigma }(%
\mathbf{r-e}_{z})+\psi _{\sigma }^{\dag }(\mathbf{r})T_{2}\psi _{\sigma }(\mathbf{r+e}_{x}-%
\mathbf{e}_{z}) \notag \\
&+&\psi _{\sigma }^{\dag }(\mathbf{r})T_{2}^{\prime }\psi
_{\sigma }(\mathbf{r-e}_{x}+\mathbf{e}_{z}),
\end{eqnarray}
where $\sigma=e, o$ stands for even and odd sites, respectively. $\mathrm{T}_{\sigma \sigma^{\prime}}$ is the  hopping matrix and $\mathbf{r}, \mathbf{r}^{\prime}$ label the location of until cell as shown in Fig.~\ref{fig:lattice}. The hopping matrices can be expressed as
\begin{eqnarray}
T_{0} &=&\left(
\begin{array}{cc}
0 & t \\
t & 0%
\end{array}%
\right) ,  \notag \\
T_{1x} &=&T_{1z}^{\prime }=T_{2}=\left(
\begin{array}{cc}
0 & t \\
0 & 0%
\end{array}%
\right) , \notag \\
T_{1x}^{\prime } &=&T_{1z}=T_{2}^{\prime }=\left(
\begin{array}{cc}
0 & 0  \\
t & 0%
\end{array}%
\right) ,
\end{eqnarray}

In the momentum space, the hopping term $H_{hop}$ can be written as
\begin{equation}
H_{hop}(\mathbf{k})=\underset{\mathbf{k}}{\sum }\left[ \psi _{e}^{\dag }(%
\mathbf{k}),\psi _{o}^{\dag }(\mathbf{k})\right] \left[
\begin{array}{cc}
0 & \epsilon _{eo}(\mathbf{k}) \\
\epsilon _{oe}(\mathbf{k}) & 0%
\end{array}%
\right] \left[
\begin{array}{c}
\psi _{e}(\mathbf{k}) \\
\psi _{o}(\mathbf{k})%
\end{array}%
\right] ,
\end{equation}
where $a$ is the lattice constant and the band dispersion $\epsilon _{eo}(\mathbf{k})=-t\{1+e^{ik_{x}\sqrt{2}a}+e^{-ik_{z}\sqrt{2}%
a}+e^{ik_{x}\sqrt{2}a-ik_{z}\sqrt{2}a}\}$, $\epsilon _{oe}(\mathbf{k})=-t\{1+e^{-ik_{x}\sqrt{2}a}+e^{ik_{z}\sqrt{2}%
a}+e^{-ik_{x}\sqrt{2}a+ik_{z}\sqrt{2}a}\}$.

\section{Local single-particle and two-particle Green's function}

In the absence of hopping, i.e., $t=0$, the local Hamiltonian can be defined from Eq.(1) as
$H_{0}=\frac{U}{2}\underset{\mathbf{r},\sigma}{\sum }n_{\sigma }(\mathbf{r}%
)(n_{\sigma }(\mathbf{r})-1)-\underset{\mathbf{r},\sigma}{\sum }\mu _{\sigma
}n_{\sigma }(\mathbf{r})-\delta _{c}\left\vert \alpha \right\vert ^{2}$ with the energy detuning $\delta _{c}$. Then, the local single-particle Green's function can be calculated in the operator representation via the occupation number basis as
\begin{equation}
G_{ee}(\tau )=-\left\langle T_{\tau }\hat{\psi}_{e}(\tau )\psi _{e}^{\dagger
}(0)\right\rangle =-\frac{1}{Z_{0}}Tr\left[ e^{-(\beta -\tau )H_{0}}\hat{\psi%
}_{e}e^{-\tau H_{0}}\psi _{e}^{\dagger }\right] , \label{B2}
\end{equation}%
where $Z_{0}= Tr e^{-\beta H_{0}}$ and $|n_{e}n_{o}\rangle =\frac{\left( \hat{\psi}%
_{e}^{\dagger }\right) ^{n_{e}}\left( \hat{\psi}_{o}^{\dagger }\right)
^{n_{o}}}{\sqrt{n_{e}!n_{o}!}}|0\rangle$ is
the eigenbasis of the local Hamiltonian $H_{0}$ with eigenvalue $\varepsilon _{n_{e},n_{o}}=\frac{U}{2}n_{o}(n_{o}-1)+\frac{U}{%
2}n_{e}(n_{e}-1)-\mu _{o}n_{o}-\mu _{e}n_{e}-\delta _{c}\left\vert \alpha
\right\vert ^{2}$.
After implementing the fourier transform, the above correlator can be expressed as
\begin{eqnarray}
G_{ee}(i\omega )=\int_{0}^{\beta }d\tau e^{i\omega \tau }G_{ee}(\tau )
=-\frac{1}{Z_{0}}\underset{n_{e},n_{o}}{\overset{\infty }{\sum }}(n_{e}+1)%
\frac{e^{-\beta \varepsilon _{n_{e}+1,n_{o}}}-e^{-\beta \varepsilon
_{n_{e},n_{o}}}}{i\omega +\varepsilon _{n_{e},n_{o}}-\varepsilon
_{n_{e}+1,n_{o}}},   \label{B4}
\end{eqnarray}
In the low temperature limit $T\rightarrow 0(\beta U \gg 1)$, the exponential term in Eq.~\eqref{B4} selects out the ground state of local Hamiltonian and contributions of other states are suppressed. Therefore, for instance, $G_{ee}$ can be calculated as
\begin{equation}
G_{ee}(i\omega )=\frac{\bar{n}_{e}+1}{i\omega +\varepsilon _{\bar{n}%
_{e},0}-\varepsilon _{\bar{n}_{e}+1,0}}-\frac{\bar{n}_{e}}{i\omega
+\varepsilon _{\bar{n}_{e}-1,0}-\varepsilon _{\bar{n}_{e},0}},  \label{B5}
\end{equation}
where $\bar{n}_{e}$ and $\bar{n}_{o}$ can be determined through minimizing the ground state energy  $\varepsilon _{\bar{n}_{e}, \bar{n}_{o}}={\min }_{n_{e},n_{o}}\varepsilon _{n_{e},n_{o}}$.
In the same way, we can also obtain the local Green's function $G_{o o}$ as
\begin{equation}
G_{o o}(i \omega)=\frac{\bar{n}_{o}+1}{i \omega+\varepsilon_{0, \bar{n}%
_{o}}-\varepsilon_{0, \bar{n}_{o}+1}}-\frac{\bar{n}_{o}}{i
\omega+\varepsilon_{0, \bar{n}_{o}-1}-\varepsilon_{0, \bar{n}_{o}}},
\label{B6}
\end{equation}

The two-particle Green's function can be calculated in the similar way as
\begin{eqnarray}
\chi _{\sigma \sigma ^{\prime }}(\tau _{1},\tau _{2},\tau _{3},0)
&=&\left\langle \psi _{\sigma }(\tau _{1})\psi _{\sigma }^{\ast }(\tau
_{2})\psi _{\sigma ^{\prime }}(\tau _{3})\psi _{\sigma ^{\prime }}^{\ast
}(0)\right\rangle _{0}^{c} \nonumber \\
&=&\frac{1}{Z_{0}}Tr\left[ e^{-\beta H_{0}}\left\langle T_{\tau }\psi
_{\sigma }(\tau _{1})\psi _{\sigma }^{\dagger }(\tau _{2})\psi _{\sigma
^{\prime }}(\tau _{3})\psi _{\sigma ^{\prime }}^{\dagger }(0)\right\rangle %
\right],   \label{B7}
\end{eqnarray}
where $\sigma =e,o$. To calculate the parameter $g_{\sigma \sigma ^{\prime
}}$ in the static limit, we only consider the time average of $\chi _{\sigma
\sigma ^{\prime }}$. Then, the diagonal part can be calculated as
\begin{eqnarray}
\bar{\chi}_{ee} &=&\int_{0}^{\beta }d\tau _{1}d\tau _{2}d\tau _{3}\chi
_{ee}\left( \tau _{1},\tau _{2},\tau _{3},0\right) -2\beta \lbrack
G_{ee}(i\omega =0)]^{2}  \notag \\
&=&\frac{-4(\bar{n}_{e}+1)(\bar{n}_{e}+2)}{\left( \varepsilon _{\bar{n}%
_{e},0}-\varepsilon _{\bar{n}_{e}+1,0}\right) ^{2}\left( \varepsilon _{\bar{n%
}_{e},0}-\varepsilon _{\bar{n}_{e}+2,0}\right) }+\frac{-4\bar{n}_{e}(\bar{n}%
_{e}-1)}{\left( \varepsilon _{\bar{n}_{e},0}-\varepsilon _{\bar{n}%
_{e}-1,0}\right) ^{2}\left( \varepsilon _{\bar{n}_{e},0}-\varepsilon _{\bar{n%
}_{e}-2,0}\right) } \notag \\
&+&\frac{-4\bar{n}_{e}(\bar{n}_{e}+1)}{\left( \varepsilon _{\bar{n}%
_{e}-1,0}-\varepsilon _{\bar{n}_{e},0}\right) ^{2}\left( \varepsilon _{\bar{n%
}_{e}+1,0}-\varepsilon _{\bar{n}_{e},0}\right) }+\frac{4\bar{n}_{e}(\bar{n}%
_{e}+1)}{\left( \varepsilon _{\bar{n}_{e},0}-\varepsilon _{\bar{n}%
_{e}+1,0}\right) ^{2}\left( \varepsilon _{\bar{n}_{e},0}-\varepsilon _{\bar{n%
}_{e}-1,0}\right) } \notag \\
&+&\frac{4(\bar{n}_{e}+1)^{2}}{\left( \varepsilon _{\bar{n}%
_{e},0}-\varepsilon _{\bar{n}_{e}+1,0}\right) ^{3}}+\frac{-4\bar{n}_{e}^{2}}{%
\left( \varepsilon _{\bar{n}_{e}-1,0}-\varepsilon _{\bar{n}_{e},0}\right)
^{3}},    \label{B10}
\end{eqnarray}
with \begin{eqnarray*}
\chi _{ee}(\tau _{1},\tau _{2},\tau _{3},0)& = &\frac{1}{Z_{0}}\overset{%
\infty }{\underset{n_{e}=0}{\sum }}e^{-\beta \varepsilon _{n_{e},0}}\{\theta
(\tau _{1}-\tau _{2})\theta (\tau _{2}-\tau _{3})e^{\tau _{1}(\varepsilon
_{n_{e},0}-\varepsilon _{n_{e}+1,0})+\tau _{2}(\varepsilon
_{n_{e}+1,0}-\varepsilon _{n_{e},0})+\tau _{3}(\varepsilon
_{n_{e},0}-\varepsilon _{n_{e}+1,0})}(n_{e}+1)^{2} \\
&+& \theta (\tau _{1}-\tau _{3})\theta (\tau _{3}-\tau _{2})e^{\tau
_{1}(\varepsilon _{n_{e},0}-\varepsilon _{n_{e}+1,0})+\tau _{2}(\varepsilon
_{n_{e}+2,0}-\varepsilon _{n_{e}+1,0})+\tau _{3}(\varepsilon
_{n_{e}+1,0}-\varepsilon _{n_{e}+2,0})}(n_{e}+1)(n_{e}+2) \\
&+&\theta (\tau _{3}-\tau _{1})\theta (\tau _{1}-\tau _{2})e^{\tau
_{1}(\varepsilon _{n_{e}+1,0}-\varepsilon _{n_{e}+2,0})+\tau
_{2}(\varepsilon _{n_{e}+2,0}-\varepsilon _{n_{e}+1,0})+\tau
_{3}(\varepsilon _{n_{e},0}-\varepsilon _{n_{e}+1,0})}(n_{e}+1)(n_{e}+2) \\
&+&\theta (\tau _{2}-\tau _{1})\theta (\tau _{1}-\tau _{3})e^{\tau
_{1}(\varepsilon _{n_{e}-1,0}-\varepsilon _{n_{e},0})+\tau _{2}(\varepsilon
_{n_{e},0}-\varepsilon _{n_{e}-1,0})+\tau _{3}(\varepsilon
_{n_{e},0}-\varepsilon _{n_{e}+1,0})}(n_{e}+1)n_{e} \\
&+&\theta (\tau _{2}-\tau _{3})\theta (\tau _{3}-\tau _{1})e^{\tau
_{1}(\varepsilon _{n_{e},0}-\varepsilon _{n_{e}+1,0})+\tau _{2}(\varepsilon
_{n_{e},0}-\varepsilon _{n_{e}-1,0})+\tau _{3}(\varepsilon
_{n_{e}-1,0}-\varepsilon _{n_{e},0})}(n_{e}+1)n_{e} \\
&+&\theta (\tau _{3}-\tau _{2})\theta (\tau _{2}-\tau _{1})e^{\tau
_{1}(\varepsilon _{n_{e},0}-\varepsilon _{n_{e}+1,0})+\tau _{2}(\varepsilon
_{n_{e}+1,0}-\varepsilon _{n_{e},0})+\tau _{3}(\varepsilon
_{n_{e},0}-\varepsilon _{n_{e}+1,0})}(n_{e}+1)^{2}\}.
\end{eqnarray*}

Similarly, other diagonal part can also be calculated as follows
\begin{eqnarray}
\bar{\chi}_{oo} &=&\int_{0}^{\beta }d\tau _{1}d\tau _{2}d\tau _{3}\chi
_{oo}\left( \tau _{1},\tau _{2},\tau _{3},0\right) -2\beta \lbrack
G_{oo}(i\omega =0)]^{2}  \notag \\
&=&\frac{-4(\bar{n}_{o}+1)(\bar{n}_{o}+2)}{\left( \varepsilon _{0,\bar{n}%
_{o}}-\varepsilon _{0,\bar{n}_{o}+1}\right) ^{2}\left( \varepsilon _{0,\bar{n%
}_{o}}-\varepsilon _{0,\bar{n}_{o}+2}\right) }+\frac{-4\bar{n}_{o}(\bar{n}%
_{o}-1)}{\left( \varepsilon _{0,\bar{n}_{o}}-\varepsilon _{0,\bar{n}%
_{o}-1}\right) ^{2}\left( \varepsilon _{0,\bar{n}_{o}}-\varepsilon _{0,\bar{n%
}_{o}-2}\right) } \notag \\
&+& \frac{-4\bar{n}_{o}(\bar{n}_{o}+1)}{\left( \varepsilon _{0,\bar{n}%
_{o}-1}-\varepsilon _{0,\bar{n}_{o}}\right) ^{2}\left( \varepsilon _{0,\bar{n%
}_{o}+1}-\varepsilon _{0,\bar{n}_{o}}\right) }+\frac{4\bar{n}_{o}(\bar{n}%
_{o}+1)}{\left( \varepsilon _{0,\bar{n}_{o}}-\varepsilon _{0,\bar{n}%
_{o}+1}\right) ^{2}\left( \varepsilon _{0,\bar{n}_{o}}-\varepsilon _{0,\bar{n%
}_{o}-1}\right) } \notag \\
&+& \frac{4(\bar{n}_{o}+1)^{2}}{\left( \varepsilon _{0,\bar{n}%
_{o}}-\varepsilon _{0,\bar{n}_{o}+1}\right) ^{3}}+\frac{-4\bar{n}_{o}^{2}}{%
\left( \varepsilon _{0,\bar{n}_{o}-1}-\varepsilon _{0,\bar{n}_{o}}\right)
^{3}},   \label{B11}
\end{eqnarray}
In the similar way, we can obtain the off-diagonal part in the static limit as
\begin{equation}
\bar{\chi}_{eo}=\bar{n}_{e}f_{1}+(\bar{n}_{e}+1)f_{2}+\bar{n}_{e}f_{3}+(\bar{%
n}_{e}+1)f_{4},  \label{B15}
\end{equation}
with
\begin{eqnarray*}
f_{1} &=&\{\frac{-\bar{n}_{o}}{\left( \varepsilon _{\bar{n}_{e},\bar{n}%
_{o}}-\varepsilon _{\bar{n}_{e},\bar{n}_{o}-1}\right) ^{2}\left( \varepsilon
_{\bar{n}_{e},\bar{n}_{o}}-\varepsilon _{\bar{n}_{e}-1,\bar{n}_{o}-1}\right)
}+\frac{-\bar{n}_{o}}{\left( \varepsilon _{\bar{n}_{e},\bar{n}%
_{o}}-\varepsilon _{\bar{n}_{e}-1,\bar{n}_{o}}\right) ^{2}\left( \varepsilon
_{\bar{n}_{e},\bar{n}_{o}}-\varepsilon _{\bar{n}_{e}-1,\bar{n}_{o}-1}\right)
}  \notag \\
&&+\frac{\bar{n}_{o}}{\left( \varepsilon _{\bar{n}_{e},\bar{n}%
_{o}}-\varepsilon _{\bar{n}_{e},\bar{n}_{o}-1}\right) ^{2}\left( \varepsilon
_{\bar{n}_{e},\bar{n}_{o}}-\varepsilon _{\bar{n}_{e}-1,\bar{n}_{o}}\right) }+%
\frac{\bar{n}_{o}}{\left( \varepsilon _{\bar{n}_{e},\bar{n}_{o}}-\varepsilon
_{\bar{n}_{e}-1,\bar{n}_{o}}\right) ^{2}\left( \varepsilon _{\bar{n}_{e},%
\bar{n}_{o}}-\varepsilon _{\bar{n}_{e},\bar{n}_{o}-1}\right) } \\
&&+\frac{-2\bar{n}_{o}}{\left( \varepsilon _{\bar{n}_{e},\bar{n}%
_{o}}-\varepsilon _{\bar{n}_{e}-1,\bar{n}_{o}}\right) \left( \varepsilon _{%
\bar{n}_{e},\bar{n}_{o}}-\varepsilon _{\bar{n}_{e},\bar{n}_{o}-1}\right)
\left( \varepsilon _{\bar{n}_{e},\bar{n}_{o}}-\varepsilon _{\bar{n}_{e}-1,%
\bar{n}_{o}-1}\right) }\}  \notag  \label{B16}
\end{eqnarray*}

\begin{eqnarray*}
f_{2} &=&\{\frac{-\bar{n}_{o}}{\left( \varepsilon _{\bar{n}_{e},\bar{n}%
_{o}}-\varepsilon _{\bar{n}_{e},\bar{n}_{o}-1}\right) ^{2}\left( \varepsilon
_{\bar{n}_{e},\bar{n}_{o}}-\varepsilon _{\bar{n}_{e}+1,\bar{n}_{o}-1}\right)
}+\frac{-\bar{n}_{o}}{\left( \varepsilon _{\bar{n}_{e},\bar{n}%
_{o}}-\varepsilon _{\bar{n}_{e}+1,\bar{n}_{o}}\right) ^{2}\left( \varepsilon
_{\bar{n}_{e},\bar{n}_{o}}-\varepsilon _{\bar{n}_{e}+1,\bar{n}_{o}-1}\right)
}  \notag \\
&&+\frac{\bar{n}_{o}}{\left( \varepsilon _{\bar{n}_{e},\bar{n}%
_{o}-1}-\varepsilon _{\bar{n}_{e},\bar{n}_{o}}\right) ^{2}\left( \varepsilon
_{\bar{n}_{e},\bar{n}_{o}}-\varepsilon _{\bar{n}_{e}+1,\bar{n}_{o}}\right) }+%
\frac{\bar{n}_{o}}{\left( \varepsilon _{\bar{n}_{e},\bar{n}_{o}}-\varepsilon
_{\bar{n}_{e}+1,\bar{n}_{o}}\right) ^{2}\left( \varepsilon _{\bar{n}_{e},%
\bar{n}_{o}}-\varepsilon _{\bar{n}_{e},\bar{n}_{o}-1}\right) } \\
&&+\frac{-2\bar{n}_{o}}{\left( \varepsilon _{\bar{n}_{e},\bar{n}%
_{o}}-\varepsilon _{\bar{n}_{e}+1,\bar{n}_{o}}\right) \left( \varepsilon _{%
\bar{n}_{e},\bar{n}_{o}}-\varepsilon _{\bar{n}_{e}+1,\bar{n}_{o}-1}\right)
\left( \varepsilon _{\bar{n}_{e},\bar{n}_{o}}-\varepsilon _{\bar{n}_{e},\bar{%
n}_{o}-1}\right) }\}  \notag  \label{B17}
\end{eqnarray*}

\begin{eqnarray*}
f_{3} &=&\{\frac{-(\bar{n}_{o}+1)}{\left( \varepsilon _{\bar{n}_{e},\bar{n}%
_{o}}-\varepsilon _{\bar{n}_{e}-1,\bar{n}_{o}}\right) ^{2}\left( \varepsilon
_{\bar{n}_{e},\bar{n}_{o}}-\varepsilon _{\bar{n}_{e}-1,\bar{n}_{o}+1}\right)
}+\frac{(\bar{n}_{o}+1)}{\left( \varepsilon _{\bar{n}_{e},\bar{n}%
_{o}}-\varepsilon _{\bar{n}_{e}-1,\bar{n}_{o}}\right) ^{2}\left( \varepsilon
_{\bar{n}_{e},\bar{n}_{o}}-\varepsilon _{\bar{n}_{e},\bar{n}_{o}+1}\right) }
\notag \\
&&+\frac{(\bar{n}_{o}+1)}{\left( \varepsilon _{\bar{n}_{e},\bar{n}%
_{o}}-\varepsilon _{\bar{n}_{e},\bar{n}_{o}+1}\right) ^{2}\left( \varepsilon
_{\bar{n}_{e},\bar{n}_{o}}-\varepsilon _{\bar{n}_{e}-1,\bar{n}_{o}}\right) }+%
\frac{-(\bar{n}_{o}+1)}{\left( \varepsilon _{\bar{n}_{e},\bar{n}%
_{o}}-\varepsilon _{\bar{n}_{e},\bar{n}_{o}+1}\right) ^{2}\left( \varepsilon
_{\bar{n}_{e},\bar{n}_{o}}-\varepsilon _{\bar{n}_{e}-1,\bar{n}_{o}+1}\right)
} \\
&&+\frac{-2(\bar{n}_{o}+1)}{\left( \varepsilon _{\bar{n}_{e},\bar{n}%
_{o}}-\varepsilon _{\bar{n}_{e}-1,\bar{n}_{o}}\right) \left( \varepsilon _{%
\bar{n}_{e},\bar{n}_{o}}-\varepsilon _{\bar{n}_{e},\bar{n}_{o}+1}\right)
\left( \varepsilon _{\bar{n}_{e},\bar{n}_{o}}-\varepsilon _{\bar{n}_{e}-1,%
\bar{n}_{o}+1}\right) }\}  \notag  \label{B18}
\end{eqnarray*}

\begin{eqnarray*}
f_{4} &=&\{\frac{(\bar{n}_{o}+1)}{\left( \varepsilon _{\bar{n}_{e},\bar{n}%
_{o}}-\varepsilon _{\bar{n}_{e},\bar{n}_{o}+1}\right) ^{2}\left( \varepsilon
_{\bar{n}_{e},\bar{n}_{o}}-\varepsilon _{\bar{n}_{e}+1,\bar{n}_{o}}\right) }+%
\frac{(\bar{n}_{o}+1)}{\left( \varepsilon _{\bar{n}_{e},\bar{n}%
_{o}}-\varepsilon _{\bar{n}_{e}+1,\bar{n}_{o}}\right) ^{2}\left( \varepsilon
_{\bar{n}_{e},\bar{n}_{o}}-\varepsilon _{\bar{n}_{e},\bar{n}_{o}+1}\right) }
\notag \\
&&+\frac{(\bar{n}_{o}+1)}{\left( \varepsilon _{\bar{n}_{e},\bar{n}%
_{o}}-\varepsilon _{\bar{n}_{e},\bar{n}_{o}+1}\right) ^{2}\left( \varepsilon
_{\bar{n}_{e},\bar{n}_{o}}-\varepsilon _{\bar{n}_{e}+1,\bar{n}_{o}+1}\right)
}+\frac{-(\bar{n}_{o}+1)}{\left( \varepsilon _{\bar{n}_{e},\bar{n}%
_{o}}-\varepsilon _{\bar{n}_{e}+1,\bar{n}_{o}}\right) ^{2}\left( \varepsilon
_{\bar{n}_{e},\bar{n}_{o}}-\varepsilon _{\bar{n}_{e}+1,\bar{n}_{o}+1}\right)
} \\
&&+\frac{-2(\bar{n}_{o}+1)}{\left( \varepsilon _{\bar{n}_{e},\bar{n}%
_{o}}-\varepsilon _{\bar{n}_{e}+1,\bar{n}_{o}}\right) \left( \varepsilon _{%
\bar{n}_{e},\bar{n}_{o}}-\varepsilon _{\bar{n}_{e}+1,\bar{n}_{o}+1}\right)
\left( \varepsilon _{\bar{n}_{e},\bar{n}_{o}}-\varepsilon _{\bar{n}_{e},\bar{%
n}_{o}+1}\right) }\}.  \notag  \label{B19}
\end{eqnarray*}
Up to this point, we have obtained the two-point and four-point correlators from
the local Hamiltonian. Then, $g_{\sigma \sigma ^{\prime }}$ in Eq. (6) of the main text can be determined by the two-particle vertex in the static limit,  which can be written as
\begin{equation}
g_{\sigma \sigma ^{\prime }}=\frac{-\bar{\chi}_{\sigma \sigma ^{\prime }}}{(%
\bar{G}_{\sigma \sigma }\bar{G}_{\sigma ^{\prime }\sigma ^{\prime }})^{2}+(%
\bar{G}_{\sigma \sigma })^{4}\delta _{\sigma \sigma ^{\prime }}},
\label{B21}
\end{equation}
where $\bar{G}_{\sigma \sigma }={G}_{\sigma \sigma }(i\omega =0)$.

\section{Double Hubbard-Stronovich transformation}

The connected correlators of the original boson fields $\psi_{\sigma}$ can be obtained from the generating function%
\begin{equation}
Z[J^{\ast},J]=\int D[\psi^{\ast},\psi]\exp\left\{  \psi_{\sigma}^{\ast
}T_{\sigma\sigma^{\prime}}\psi_{\sigma^{\prime}}-S_{0}[\psi_{\sigma}^{\ast
},\psi_{\sigma}]+[(J|\psi)+c.c]\right\}  ,  \label{C1}
\end{equation}
where $J,J^{\ast}$ are external sources and  $(J|\psi)=\sum_{\sigma, \mathrm{r}} \int_{0}^{\beta} d \tau J_{\sigma}^{*}(\mathbf{r}) \psi_{\sigma}(\mathbf{r})$. Introducing the first Hubbard-Stratonovich transformation with auxiliary field $\phi$,
\begin{equation}
Z[J^{\ast},J]=\int D[\psi^{\ast},\psi;\phi^{\ast},\phi]\exp\left\{
-\phi_{\sigma}^{\ast}T_{\sigma\sigma^{\prime}}^{-1}\phi_{\sigma^{\prime}%
}-S_{0}[\psi_{\sigma}^{\ast},\psi_{\sigma}]+[(\phi|\psi)+c.c]+[(J|\psi
)+c.c]\right\}  ,  \label{C2}
\end{equation}
After a shift $\phi_{\sigma}\longrightarrow\phi_{\sigma}-J_{\sigma},\phi_{%
\sigma}^{\ast}\longrightarrow\phi_{\sigma}^{\ast}-J_{\sigma}^{\ast}$, we obtain
\begin{equation}
Z[J^{\ast},J]=\int D[\psi^{\ast},\psi;\phi^{\ast},\phi]\exp\left\{  -\left(
\phi_{\sigma}^{\ast}-J_{\sigma}^{\ast}\right)  T_{\sigma\sigma^{\prime}}%
^{-1}\left(  \phi_{\sigma^{\prime}}-J_{\sigma^{\prime}}\right)  -S_{0}%
[\psi_{\sigma}^{\ast},\psi_{\sigma}]+[(\phi|\psi)+c.c]\right\}  , \label{C3}
\end{equation}
Integrating $\psi_{\sigma}$ fields, we then obtain
\begin{equation}
Z[J^{\ast},J]=Z_{0}\int D[\phi^{\ast},\phi]\exp\left\{  -\left(  \phi_{\sigma
}^{\ast}-J_{\sigma}^{\ast}\right)  T_{\sigma\sigma^{\prime}}^{-1}\left(
\phi_{\sigma^{\prime}}-J_{\sigma^{\prime}}\right)  +W[\phi_{\sigma}^{\ast
},\phi_{\sigma}]\right\}  ,  \label{C4}
\end{equation}
Next, applying the second Hubbard-Stratonovich transformation with auxiliary field $\psi^{\prime}$,
we obtain
\begin{eqnarray}
\begin{split}
Z[J^{\ast},J] &  =Z_{0}\int D[\phi^{\ast},\phi;\psi^{\prime\ast},\psi^{\prime
}]\exp\left\{  \psi_{\sigma}^{\prime\ast}T_{\sigma\sigma^{\prime}}\psi
_{\sigma}^{\prime}-[(\psi^{\prime}|\phi-J)+c.c]+W[\phi_{\sigma}^{\ast}%
,\phi_{\sigma}]\right\}  \\
&  =Z_{0}\int D[\phi^{\ast},\phi;\psi^{\prime\ast},\psi^{\prime}]\exp\left\{
\psi_{\sigma}^{\prime\ast}T_{\sigma\sigma^{\prime}}\psi_{\sigma}^{\prime
}-[(\psi^{\prime}|\phi)+c.c]+[(\psi^{\prime}|J)+c.c]+W[\phi_{\sigma}^{\ast
},\phi_{\sigma}]\right\}  , \label{C5}
\end{split}
\end{eqnarray}
After integrating $\phi_{\sigma}$ fields, we get
\begin{equation}
Z[J^{\ast},J]=Z_{0}\int D[\psi^{\prime\ast},\psi^{\prime
}]\exp\left\{  \psi_{\sigma}^{\prime\ast}T_{\sigma\sigma^{\prime}}\psi
_{\sigma}^{\prime}+W[\psi_{\sigma}^{\prime\ast},\psi_{\sigma}^{\prime}%
]+[(\psi^{\prime}|J)+c.c]\right\}.  \label{C6}
\end{equation}
From the above equation, we prove that $Z[J^{\ast},J]$ is also the generating function of $\psi_{\sigma }^{\prime}$ field, which is equal to the generating
function of orginal $\psi$ field. This proves that the connected correlators of $\psi_{\sigma}^{\prime}$ are
the same as that of $\psi_{\sigma}$. Therefore, $\psi^{\prime}_{\sigma}$ is identified with the original boson field $\psi_{\sigma}$. We can thus use the same notation for both fields.

\section{Coefficients of the effective action at the critical region}

To obtain the effective action at the critical region, we first rewrite the
action Eq.(6) in the main text in the momentum space as
\begin{eqnarray}
S_{k} &=&\underset{\mathbf{k},\omega}{\sum }\psi _{e}^{\ast }(\mathbf{k,}i\omega
)\left[ -G_{ee}^{-1}(i\omega)\right] \psi _{e}(\mathbf{k,}i\omega
)+\psi _{o}^{\ast }(\mathbf{k,}i\omega)\left[ -G_{oo}^{-1}(i\omega
)\right] \psi _{o}(\mathbf{k,} i\omega)  \notag \\
&+&\psi _{e}^{\ast }(\mathbf{k,}i\omega )\epsilon _{eo}(\mathbf{k}%
) \psi _{o}(\mathbf{k,}i\omega )+\psi _{o}^{\ast }(\mathbf{k,}%
i\omega )\epsilon _{oe}(\mathbf{k}) \psi _{e}(\mathbf{k,}%
i\omega )+\frac{1}{2}g_{ee}\left\vert \psi _{e}(\mathbf{k,}i\omega)\right\vert
^{4} \nonumber \\
&+&\frac{1}{2}g_{oo}\left\vert \psi _{o}(\mathbf{k,}i\omega
)\right\vert ^{4}+\frac{1}{2}g_{eo}\left\vert \psi _{e}(\mathbf{k,}%
i\omega)\right\vert ^{2}\left\vert \psi _{o}(\mathbf{k,}i\omega)\right\vert ^{2}+\frac{1}{2}g_{oe}\left\vert \psi _{o}(\mathbf{k,}%
i\omega)\right\vert ^{2}\left\vert \psi _{e}(\mathbf{k,}i\omega)\right\vert ^{2},
\end{eqnarray}
Next, we expand the inverse local single particle Green's function $%
G_{\sigma \sigma }^{-1}(i\omega)$ and dispersion $\epsilon _{\sigma
\sigma ^{\prime }}(\mathbf{k})$ to the quadratic order in $\mathbf{k}$ and
$\omega$, respectively and obtain
\begin{eqnarray}
S_{k} &=&\underset{\mathbf{k},\omega}{\sum }r_{e}\psi _{e}^{\ast }(\mathbf{k,}%
i\omega)\psi _{e}(\mathbf{k,}i\omega)+K_{e1}\psi _{e}^{\ast }(%
\mathbf{k,}i\omega)(-i\omega)\psi _{e}(\mathbf{k,}i\omega
)+K_{e2}(i\omega)\psi _{e}^{\ast }(\mathbf{k,}i\omega
)(-i\omega)\psi _{e}(\mathbf{k,}i\omega)  \notag \notag \\
&+&r_{o}\psi _{o}^{\ast }(\mathbf{k,}i\omega)\psi _{o}(\mathbf{k,}%
i\omega)+K_{o1}\psi _{o}^{\ast }(\mathbf{k,}i\omega)(-i\omega)\psi _{o}(\mathbf{k,}i\omega)+K_{o2}(i\omega)\psi
_{o}^{\ast }(\mathbf{k,}i\omega)(-i\omega)\psi _{o}(\mathbf{k,}%
i\omega) \notag \\
&+&\psi _{e}^{\ast }(\mathbf{k,}i\omega)(-4t) \psi _{o}(%
\mathbf{k,}i\omega)+K_{3}(-i\mathbf{k})\psi _{e}^{\ast }(\mathbf{k,}%
i\omega)(i\mathbf{k})\psi _{o}(\mathbf{k,}i\omega) \notag  \\
&+&\psi _{o}^{\ast }(\mathbf{k,}i\omega)(-4t) \psi _{e}(%
\mathbf{k,}i\omega)+K_{3}(-i\mathbf{k})\psi _{o}^{\ast }(\mathbf{k,}%
i\omega)(i\mathbf{k})\psi _{e}(\mathbf{k,}i\omega) \notag \\
&+&\frac{1}{2}g_{ee}\left\vert \psi _{e}(\mathbf{k,}i\omega)\right\vert
^{4}+\frac{1}{2}g_{oo}\left\vert \psi _{o}(\mathbf{k,}i\omega)\right\vert ^{4}  \notag \\
&+&\frac{1}{2}g_{eo}\left\vert \psi _{e}(\mathbf{k,}i\omega)\right\vert
^{2}\left\vert \psi _{o}(\mathbf{k,}i\omega)\right\vert ^{2}+\frac{1}{2}%
g_{oe}\left\vert \psi _{o}(\mathbf{k,}i\omega)\right\vert
^{2}\left\vert \psi _{e}(\mathbf{k,}i\omega)\right\vert ^{2},
\end{eqnarray}
where $r_{e}=-\left( \frac{\bar{n}_{e}+1}{%
\varepsilon _{\bar{n}_{e},0}-\varepsilon _{\bar{n}_{e}+1,0}}-\frac{\bar{n}%
_{e}}{\varepsilon _{\bar{n}_{e}-1,0}-\varepsilon _{\bar{n}_{e},0}}\right)
^{-1}$, $r_{o}=-\left( \frac{\bar{n}_{o}+1}{%
\varepsilon _{0,\bar{n}_{o}}-\varepsilon _{0,\bar{n}_{o}+1}}-\frac{\bar{n}%
_{o}}{\varepsilon _{0,\bar{n}_{o}-1}-\varepsilon _{0,\bar{n}_{o}}}\right)
^{-1}$, $K_{e1}=\frac{\mu
_{e}^{2}-\left( \bar{n}_{e}^{2}+\bar{n}_{e}-1\right) U^{2}+2\mu _{e}U}{\left( \mu
_{e}+U\right) ^{2}}$,
$K_{o1}=\frac{\mu
_{o}^{2}-\left(\bar{ n}_{o}^{2}+\bar{n}_{o}-1\right) U^{2}+2\mu _{o}U}{\left( \mu
_{o}+U\right) ^{2}}$, $K_{e2}=\frac{\bar{n}_{e}(\bar{n}_{e}+1)U^{2}}{%
\left( \mu _{e}+U\right) ^{3}}$,  $K_{o2}=\frac{\bar{n}_{o}(\bar{n}_{o}+1)U^{2}}{%
\left( \mu _{o}+U\right) ^{3}}$ and $K_{3}=2ta^{2}$.
Then, we can obtain the effective action Eq.(10) in the main text by transforming the action Eq.(S21) back to the real space.

\section{the lowest single particle excitation}
To find out the single particle excitation of the system,  we analytically solve the equation $det[M(\mathbf{k},i\omega)]=0$. Therefore, the single particle excitation spectrum $E(\mathbf{k})$ can be determined through the following relation
\begin{equation}
\left( C_{9}k^{4}+C_{8}k^{2}+C_{7}\right) E^{4}+\left(
C_{6}k^{4}+C_{5}k^{2}+C_{4}\right) E^{2}+\left(
C_{3}k^{4}+C_{2}k^{2}+C_{1}\right) =0,
\end{equation}
where the coefficients are defined as $C_{1} =\left(4t\right) ^{4}-2\times\left( 4t\right) ^{2}\left[ D_{1}D_{2}+D_{3}D_{4}\right] +\left[
(D_{1}^{2}-D_{3}^{2})(D_{2}^{2}-D_{4}^{2})\right]$, $C_{2} =-4\times(4t)^{3}K_{3}+4\times(4t)K_{3}\left[ D_{1}D_{2}+D_{3}D_{4}\right]$,
$C_{3}=6\times(4t)^{2}K_{3}^{2}-2K_{3}^{2}\left[ D_{1}D_{2}+D_{3}D_{4}\right]$,
$C_{4}=2\times(4t)^{2}\left[ K_{e2}D_{2}+K_{o2}D_{1}-K_{e1}K_{o1}\right]
-K_{e1}^{2}(D_{2}^{2}-D_{4}^{2})-K_{o1}^{2}(D_{1}^{2}-D_{3}^{2})
-2K_{e2}\left[ D_{1}(D_{2}^{2}-D_{4}^{2})\right] -2K_{o2}\left[
D_{2}(D_{1}^{2}-D_{3}^{2})\right]$,
$C_{5} =4\times(4t)K_{3}K_{e2}D_{2}+4\times(4t)K_{3}K_{o2}D_{1}-4\times(4t)K_{3}K_{e1}K_{o1}$,
$C_{6}=-2K_{3}^{2}K_{e1}K_{o1}+2K_{3}^{2}K_{e2}D_{2}+2K_{3}^{2}K_{o2}D_{1}$, $C_{7}
=-2\times(4t)^{2}K_{e2}K_{o2}+K_{e1}^{2}K_{o1}^{2}+2K_{o1}^{2}K_{e2}D_{1}+2K_{e1}^{2}K_{o2}D_{2}+4K_{e2}K_{o2}D_{1}D_{2}
+K_{e2}^{2}\left( D_{2}^{2}-D_{4}^{2}\right) +K_{o2}^{2}\left(
D_{1}^{2}-D_{3}^{2}\right)$, $C_{8} =4\times(4t)K_{3}K_{e2}K_{o2}$,
and $C_{9}=-2K_{3}^{2}K_{e2}K_{o2}$ with
$D_{1}=r_{e}+2g_{ee}n_{s,e},D_{2}=r_{o}+2g_{oo}n_{s,o}$, $D_{3}=g_{ee}n_{s,e}$, $D_{4}=g_{oo}n_{s,o}$. By analytically solving the above equation and expanding the energy dispersion to the quartic order of the momentum,  the lowest single particle excitation (positive branch) can be approximately determined in the long-wave limit as
\begin{eqnarray}
E^{+}_{Goldstone}\simeq\sqrt{I_{2}}k+\sqrt{I_{4}}k^{2}
\end{eqnarray}
where $I_{2} =\frac{C_{1}C_{5}}{C_{4}^{2}}-\frac{C_{2}}{C_{4}}$,
$I_{4}=-\frac{C_{1}C_{5}^{2}}{C_{4}^{3}}+\frac{C_{2}C_{5}+C_{1}C_{6}}{%
C_{4}^{2}}-\frac{C_{3}}{C_{4}}$.

\end{document}